\documentclass[preprint,showkeys,amsmath,amssymb,pra]{revtex4-1}

\usepackage{bm}
\usepackage{bbm}
\usepackage{color}
\usepackage{graphicx}   
\newcommand{\re}{{\mathrm e}}
\newcommand{\ri}{{\mathrm i}} 
\newcommand{\kB}{k_{\rm B}} 

\newcommand{\wo}{\widetilde\omega}                
\begin{document}

\title[Floquet-state cooling]
      {Floquet-state cooling} 

\author{Onno R. Diermann} 
\author{Martin Holthaus} \email{martin.holthaus@uol.de}

\affiliation{Institut f\"ur Physik, Carl von Ossietzky Universit\"at, 
	D-26111 Oldenburg, Germany}
                  
\date{October 27, 2019}

\begin{abstract}
We demonstrate that a periodically driven quantum system can adopt a 
quasistationary state which is effectively much colder than a thermal 
reservoir it is coupled to, in the sense that certain Floquet states of the 
driven-dissipative system can carry much higher population than the ground 
state of the corresponding undriven system in thermal equilibrium. This is 
made possible by a rich Fourier spectrum of the system's Floquet transition 
matrix elements, the components of which are addressed individually by a 
suitably peaked reservoir density of states. The effect is expected to be 
important for driven solid-state systems interacting with a phonon bath 
predominantly at well-defined frequencies. 
\end{abstract} 


\maketitle 

%%%%%%%%%%%%%%%%%%%%%%%%%%%%%%%%%%%%%%%%%%%%%%%%%%%%%%%%%%%%%%%%%%%%%%%%%%%%%%%%

\section{Objective}
%%%%%%%%%%%%%%%%%%%

In countless situations of basic experimental and theo\-retical interest, 
ranging from superconducting charge pumps~\cite{GasparinettiEtAl13} over 
Dirac fermions in graphene coupled to acoustic phonons~\cite{IadecolaEtAl13} 
and quantum-dot devices~\cite{StaceEtAl13} to superconductors under 
phonon driving~\cite{MurakamiEtAl17}, emitters in laser-driven 
cavities~\cite{PagelFehske17}, or few-level systems coupled to transmission 
lines~\cite{ReimerEtAl18}, one encounters periodically driven quantum systems 
interacting with their environment~\cite{BlumelEtAl91,GrahamHuebner94,Kohn01,
HoneEtAl09,ShiraiEtAl15,Liu15,IadecolaEtAl15,SeetharamEtAl15,DaiEtAl16,
VajnaEtAl16,RestrepoEtAl16,LazaridesMoessner17,TuorilaEtAl17,HartmannEtAl17,
VorbergEtAl13,SchnellEtAl18}. Such systems usually adopt a quasistationary 
state which may depend substantially on that interaction, even to the extent 
that the dynamics are fully environment-governed~\cite{GasparinettiEtAl13}. 
An investigation of a periodically driven dissipative Bose-Hubbard model has 
demonstrated that the interaction of a driven system with a reservoir can 
protect the system against heating~\cite{IwahoriKawakami17}. In the same 
spirit, a recent study of steady states of interacting Floquet insulators 
has confirmed that a heat bath can stabilize certain low-entropy states of 
periodically driven interacting systems~\cite{SeetharamEtAl19}. This is what 
one may expect on intuitive grounds, since the system can dump the energy 
absorbed from the drive into the reservoir~\cite{LangemeyerHolthaus14,
BulnesCuetaraEtAl15}. With the present contribution we direct this line 
of research into previously unknown territory, and establish a fairly 
counter\-intuitive phenomenon: A periodically driven quantum system interacting 
with a thermal bath can effectively be made even  {\em colder\/} than the bath 
it is coupled to, ``effectively'' meaning here that a Floquet state of a 
driven-dissipative quantum system can carry much {\em higher\/} population 
than the undriven system's ground state in thermal equilibrium.

\section{Model system}
%%%%%%%%%%%%%%%%%%%%%%   
   
To demonstrate the feasibility of such unexpected Floquet-state cooling we 
employ the model of a harmonic oscillator with a periodically time-dependent 
spring function~$k(t)$, which often has served as a workhorse in studies of 
quantum thermodynamics~\cite{KohlerEtAl97,OchoaEtAl18,FreitasPaz18,
DiermannEtAl19}. It is given by the Hamiltonian
\begin{equation}
	H_0(t) = \frac{p^2}{2M} + \frac{1}{2} k(t) x^2 \; , 
\label{eq:HAM}
\end{equation}
with $M$ denoting the mass of the oscillator particle. We take the spring
function to be of the form 
\begin{equation}
	k(t) = M \Omega_0^2 - M \Omega_1^2 \cos(\omega t) \; ,  
\label{eq:MSF}
\end{equation}
so that $\Omega_0$ is the angular frequency of the undriven oscillator, and 
the frequency~$\Omega_1$ quantifies the driving strength; with this choice the 
classical equation of motion $M\ddot\xi(t) = -k(t) \,\xi(t)$ becomes equal 
to the famous Mathieu equation~\cite{AbramowitzStegun70,MagnusWinkler04}. 
Provided the parameters $\Omega_0/\omega$ and $\Omega_1/\omega$ are chosen 
such that the Mathieu equation admits stable solutions, the time-dependent 
Schr\"odinger equation with the Hamiltonian~(\ref{eq:HAM}) possesses a complete
set of square-integrable Floquet states, that is, of solutions   
\begin{equation}
	\psi_n(x,t) = u_n(x,t) \exp(-\ri\varepsilon_n t/\hbar) 
\label{eq:FLS}
\end{equation}
with square-integrable Floquet functions $u_n(x,t)$ which acquire the period 
$T = 2\pi/\omega$ of the drive,
\begin{equation}
	u_n(x,t) = u_n(x,t+T) \; .
\end{equation}	
These Floquet states~(\ref{eq:FLS}) have been obtained independently by 
several authors~\cite{PopovPerelomov70,Combescure86,Brown91}, and been 
utilized, {\em e.g.\/}, to describe the quantum dynamics of particles in a 
Paul trap~\cite{Paul90}. The key element entering into their construction 
are stable classical Floquet solutions 
\begin{equation}
	\xi(t) = v(t)\exp(\ri \nu t) 
\label{eq:CFS}	
\end{equation}
with a $T$-periodic function $v(t) = v(t+T)$ and a corresponding characteristic
exponent~$\nu$~\cite{AbramowitzStegun70,MagnusWinkler04}. The latter can be
chosen such that it connects continuously to the oscillator 
frequency~$\Omega_0$ when the drive is switched off, so that $\Omega_1$ goes 
to zero~\cite{DiermannEtAl19}. The quasienergies~$\varepsilon_n$ which
accompany the time evolution of the Floquet states~(\ref{eq:FLS}) then take
the form 
\begin{equation}  
	\varepsilon_n = \hbar\nu(n + 1/2)
	\qquad \bmod \; \hbar\omega \; , 
\label{eq:QES}
\end{equation}
where $n = 0, 1,2,3, \ldots$ is the usual integer oscillator quantum number. 
If the combination of parameters $\Omega_0/\omega$ and $\Omega_1/\omega$ 
gives rise to unstable solutions of the Mathieu equation, the quasienergy 
spectrum of the parametrically driven oscillator becomes absolutely 
continuous~\cite{Howland92} so that the system can absorb an infinite amount 
of energy from the drive; this case is not being considered here. 

Now let us couple this system~(\ref{eq:HAM}) to an infinite phonon bath 
modeled by thermally occupied harmonic oscillators with frequencies~$\wo$.
To this end, we adopt the interaction Hamiltonian~\cite{BreuerPetruccione02}
\begin{equation}
	H_{\rm int} = \gamma x \sum_{\wo} 
	\left( b^{\phantom\dagger}_{\wo} + b^\dagger_{\wo} \right) \; , 
\label{eq:SBC}
\end{equation}	
where the constant $\gamma$ carries the dimension of energy per length, and 
the operators $b^{\phantom\dagger}_{\wo}$ and $b^\dagger_{\wo}$ describe,
respectively, annihilation and creation processes in the bath. Following the 
approach pioneered by Breuer {\em et al.\/}~\cite{BreuerEtAl00}, the 
rate~$\Gamma_{fi}$ of bath-induced transitions from an initial Floquet 
state~$i$ to a final one~$f$ then is obtained as a sum 
\begin{equation}
	\Gamma_{fi} = \sum_\ell \Gamma_{fi}^{(\ell)} \; , 
\label{eq:TOR}
\end{equation}
where the partial rates
\begin{equation}
	\Gamma_{fi}^{(\ell)} = 
	\frac{2\pi}{\hbar^2} \left| V_{fi}^{(\ell)} \right|^2 
	N(\omega_{fi}^{(\ell)}) \, J(|\omega_{fi}^{(\ell)}|) 
\label{eq:PAR}
\end{equation}
correspond to the individual Floquet transition frequencies 
\begin{equation}
	\omega_{fi}^{(\ell)} = 
	(\varepsilon_f - \varepsilon_i)/\hbar + \ell\omega  
\label{eq:FTF}
\end{equation}
with $\ell = 0,\pm 1, \pm 2, \ldots \;$. The quantities $V_{fi}^{(\ell)}$ are
given by the Fourier components of the system's transition matrix elements,
\begin{equation}
	\langle u_f(t) | \, \gamma x \, | u_i(t) \rangle =
	\sum_{\ell } \re^{\ri \ell \omega t} \, V_{fi}^{(\ell)} \; ,  
\label{eq:FTM}
\end{equation}
and the numbers $N(\wo)$ specify the thermal occupation of the phonon
modes,
\begin{equation}
	N(\wo) = \left\{ \begin{array}{ll}
	  \displaystyle		
	  \frac{1}{\exp(\beta\hbar\wo) - 1} &; \quad \wo > 0 \\
	  N(-\wo) + 1 &; \quad \wo < 0 
	  \phantom{\displaystyle\int^a}
	\end{array} \right.
\label{eq:OPM}
\end{equation}					
with $\beta = 1/(\kB T_{\rm bath})$ encoding the inverse of the bath
temperature $T_{\rm bath}$, invoking the Boltzmann constant~$\kB$. Observe that
negative transition frequencies~(\ref{eq:FTF}) correspond to processes during 
which the system looses energy to the bath so that a bath phonon is created,
explaining the ``$+1$'' in the second case~(\ref{eq:OPM}) which will turn 
out to be crucial. The third input required for evaluating the partial 
rates~(\ref{eq:PAR}) is the spectral density~$J(\wo)$ of the oscillator bath; 
observe that the transition frequencies~(\ref{eq:FTF}) enter into this density 
with their absolute value only.   

Knowing the total rates~(\ref{eq:TOR}), the quasistationary distribution
$\{ p_n \}_{n = 0,1,2,\ldots}$ of the Floquet-state occupation probabilities 
characterizing the steady state is obtained as solution to the master 
equation~\cite{BreuerEtAl00}
\begin{equation}
	0 = \sum_m \big( \Gamma_{nm}p_m - \Gamma_{mn} p_n \big) \; .
\end{equation}	 			 					
For the system~(\ref{eq:HAM}) with system-bath coupling~(\ref{eq:SBC}) 
the matrix~$\Gamma$ becomes tridiagonal, connecting neighboring Floquet 
states $m = n \pm 1$ only. Moreover, for each such transition the ratio~$r$ 
of the  ``upward'' rate~$\Gamma_{n+1,n}$ to the matching ``downward'' 
rate~$\Gamma_{n,n+1}$, namely,  
\begin{equation}
	\frac{\Gamma_{n+1,n}}{\Gamma_{n,n+1}} = r 
\end{equation}
becomes independent of the oscillator quantum number~$n$. Therefore, the
Floquet-state occupation probabilities for the quasistationary state are 
given by the geometric distribution 
\begin{equation}
	p_n = (1 - r) \, r^n \; , 
\end{equation}
provided $r < 1$. If $r > 1$ the system does not reach a steady state, but 
keeps on climbing the oscillator ladder, its ``upward'' transitions being 
favored over the ``downward'' ones. Since the system's 
quasienergies~(\ref{eq:QES}) are equidistant, one may introduce a 
quasitemperature~$\tau$ by regarding~$r$ as a Boltzmann factor, 
\begin{equation}
	r = \exp\!\left(-\frac{\hbar\nu}{\kB\tau}\right) \; .   
\label{eq:DQT}
\end{equation}
Positive quasitemperatures then characterize a steady state with $r < 1$; 
the smaller~$r$, the lower~$\tau$. In contrast, negative~$\tau$ signal 
quasithermal instability. Needless to say, we are considering a nonequilibrium
system which does not possess a temperature in the sense of equilibrium
thermodynamics; the quasitemperature simply serves as a convenient parameter
for comparing the driving-engineered quasistationary state to the state that 
would be adopted in thermal equilibrium.

Now the evaluation of the general expression~(\ref{eq:PAR}) leads to the 
explicit result~\cite{DiermannEtAl19}  
\begin{equation}
	r = \frac{\displaystyle{\sum_\ell} \left| v^{(\ell)} \right|^2 \, 
		N(+\nu + \ell\omega) \; J(|\nu + \ell\omega |)}
	         {\displaystyle{\sum_\ell} \left| v^{(\ell)} \right|^2 \, 
		N(-\nu - \ell\omega) \; J(|\nu + \ell\omega |)} \; ,  
\label{eq:RAT}
\end{equation}
where $v^{(\ell)}$ denote the Fourier coefficients of the periodic parts of 
the classical Floquet solutions~(\ref{eq:CFS}),
\begin{equation}
	v(t) = \sum_\ell \re^{\ri\ell\omega t} \, v^{(\ell)} \; .   
\end{equation}
This representation~(\ref{eq:RAT}) contains the heart of the Floquet-state 
cooling mechanism. Let us assume that the spectral density $J(\wo)$ is 
particularly large at an upward transition with {\em positive\/} frequency 
$\wo = \nu + \ell_1\omega$ accompanied by a reasonably large squared Fourier 
coefficient $\left| v^{(\ell_1)} \right|^2$, but relatively small at all 
others, so that all contributions to the ratio~(\ref{eq:RAT}) with 
$\ell \neq \ell_1$ may be neglected. In this case one has approximately
\begin{equation}
	r \approx \frac{N(\nu + \ell_1\omega)}{N(\nu + \ell_1\omega) + 1} 
	= \exp\!\big[ - \beta\hbar(\nu + \ell_1\omega)\big] < 1  
\label{eq:APR}
\end{equation} 
by virtue of Eq.~(\ref{eq:OPM}). The larger $\ell_1$ can be made, that is, 
the more sizeable Fourier coefficients are available, the smaller~$r$ can be 
reached. It needs to be stressed that both the Fourier coefficient labeled 
$\ell_1$ and the density of states drop out here. These quantities set the 
scale of the corresponding partial rate~(\ref{eq:PAR}) and, hence, determine 
the time required for relaxing to the quasithermal steady state; if the 
Fourier coefficient picked out by the density of states should be small, 
this relaxation time may be quite long. Quite intriguingly, the geometric 
Floquet-state distribution implied by this approximate identity~(\ref{eq:APR}) 
looks as if the {\em driven nonequilibrium\/} system were mapped to an 
{\em undriven equilibrium\/} system characterized by a Boltzmann distribution 
with the actual temperature of the ambient bath, but with a ``renormalized'' 
effective level spacing $\hbar(\nu + \ell_1\omega)$ selected by the density 
of states, although this density itself does not figure in the end. It will 
be interesting to explore whether this particular feature exhibited by the 
present model is capable of generalization.   

Here we stick to the idea of characterizing the quasistationary steady state 
of the driven system in terms of the quasitemperature introduced through 
Eq.~(\ref{eq:DQT}). Evidently, the approximation~(\ref{eq:APR}) allows one to 
cover practically the entire interval $0 < r < 1$, implying that the range of 
quasitemperatures accessible to the system is $0 < \tau/T_{\rm bath} < \infty$.
Thus, the quasitemperature may be quite different from the bath temperature 
$T_{\rm bath}$; in particular, the driven system can effectively be much colder
than its environment.

\section{Results}
%%%%%%%%%%%%%%%%%
 
In order to substantiate this key issue we now specify a Gaussian spectral
density  
\begin{equation}
	J(\widetilde\omega) = 
	J_0 \exp\!\left(-\frac{(\wo - \wo_0)^2}{(\Delta\wo)^2}\right) 
\label{eq:GSD}
\end{equation}
centered around a frequency $\wo_0$ with width $\Delta\wo$. The parameters 
entering into the Mathieu spring function~(\ref{eq:MSF}) are chosen as 
$\Omega_0/\omega = \sqrt{2}$ and $\Omega_1/\omega = 1.0$, giving the 
characteristic exponent $\nu/\omega \approx 1.387$, only slightly 
down-shifted against the unperturbed oscillator frequency by the ac Stark 
effect~\cite{DiermannEtAl19}. Finally, the bath temperature is adjusted to 
$\hbar\omega/(\kB T_{\rm bath}) = \beta\hbar\omega = 0.1$.

\begin{figure}[t]
\centering
\includegraphics[width=0.9\linewidth]{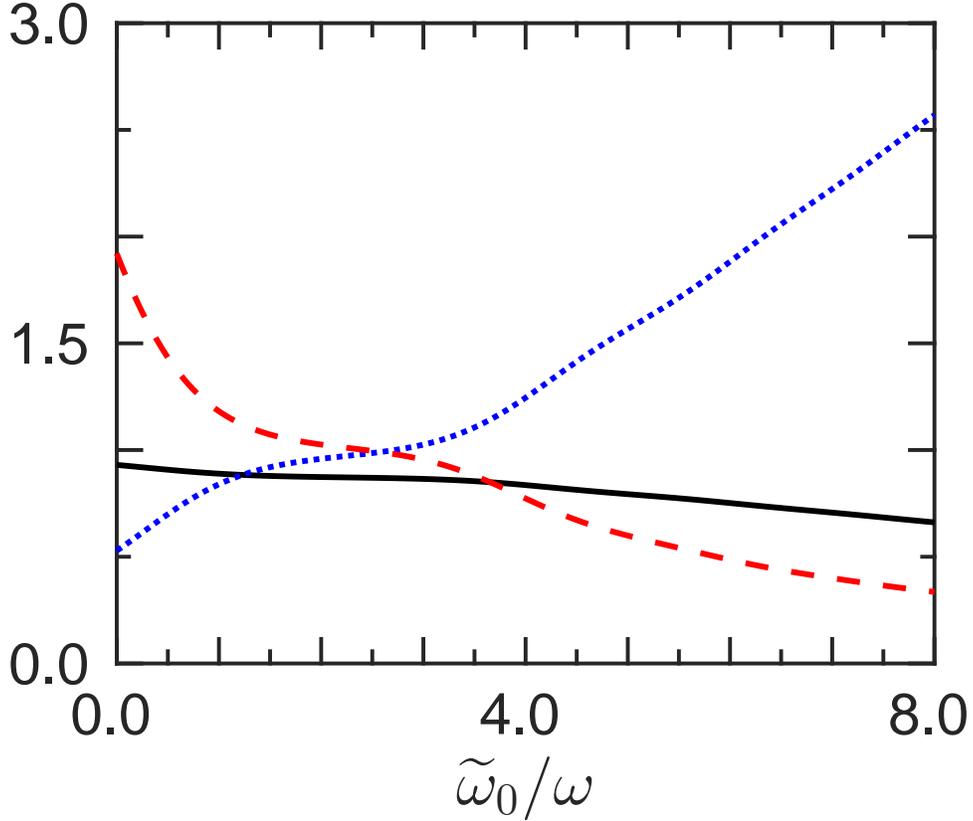}
\caption{Ratio~$r$ (full line; black), scaled quasi\-temperature
	$\tau/T_{\rm bath}$ (dashed line; red), and scaled occupation
	probability $p_0/P_0$ (dotted line; blue) of the Floquet state 
	$n = 0$ vs.\ the center $\wo_0/\omega$ of a Gaussian spectral 
	density~(\ref{eq:GSD}) with large width $\Delta\wo/\omega = 1.0$.
	The system parameters are $\Omega_0/\omega = \sqrt{2}$ and 
	$\Omega_1/\omega = 1.0$; the bath temperature has been set to 
	$\beta\hbar\omega = 0.1$.}   
\label{F_1}
\end{figure}

In Figure~\ref{F_1} we display the ratio~$r$, the scaled quasi\-temperature
$\tau/T_{\rm bath}$, and the scaled population~$p_0/P_0$ of the Floquet state 
$n = 0$, where $P_0 = 1 - \exp(-\beta\hbar\Omega_0)$ is the thermal occupation 
of the oscillator ground state without periodic driving, attained when 
$\Omega_1/\omega = 0$. Here we have chosen the width $\Delta\wo/\omega = 1.0$ 
of the spectral density~(\ref{eq:GSD}); data are plotted vs.\ its 
center~$\wo_0/\omega$. We observe a steady decrease of~$r$ with increasing 
center frequency~$\wo_0/\omega$, accompanied by the corresponding decrease of 
the quasi\-temperature, and a fairly significant increase of the population 
of the Floquet state $n = 0$, such that the latter exceeds the thermal 
equilibrium value by a factor of about~$2.5$ for $\wo_0/\omega \approx 8$. 
This finding already provides an encouraging verification of the mechanism 
underlying the approximation~(\ref{eq:APR}): Here the density~(\ref{eq:GSD}) 
successively favors Floquet transition frequencies with 
$\ell_1 = -1,0,1,2,\ldots\;$, but the Gaussian is still so wide that these 
transitions are not individually resolved.

\begin{figure}[t]
\centering
\includegraphics[width=0.9\linewidth]{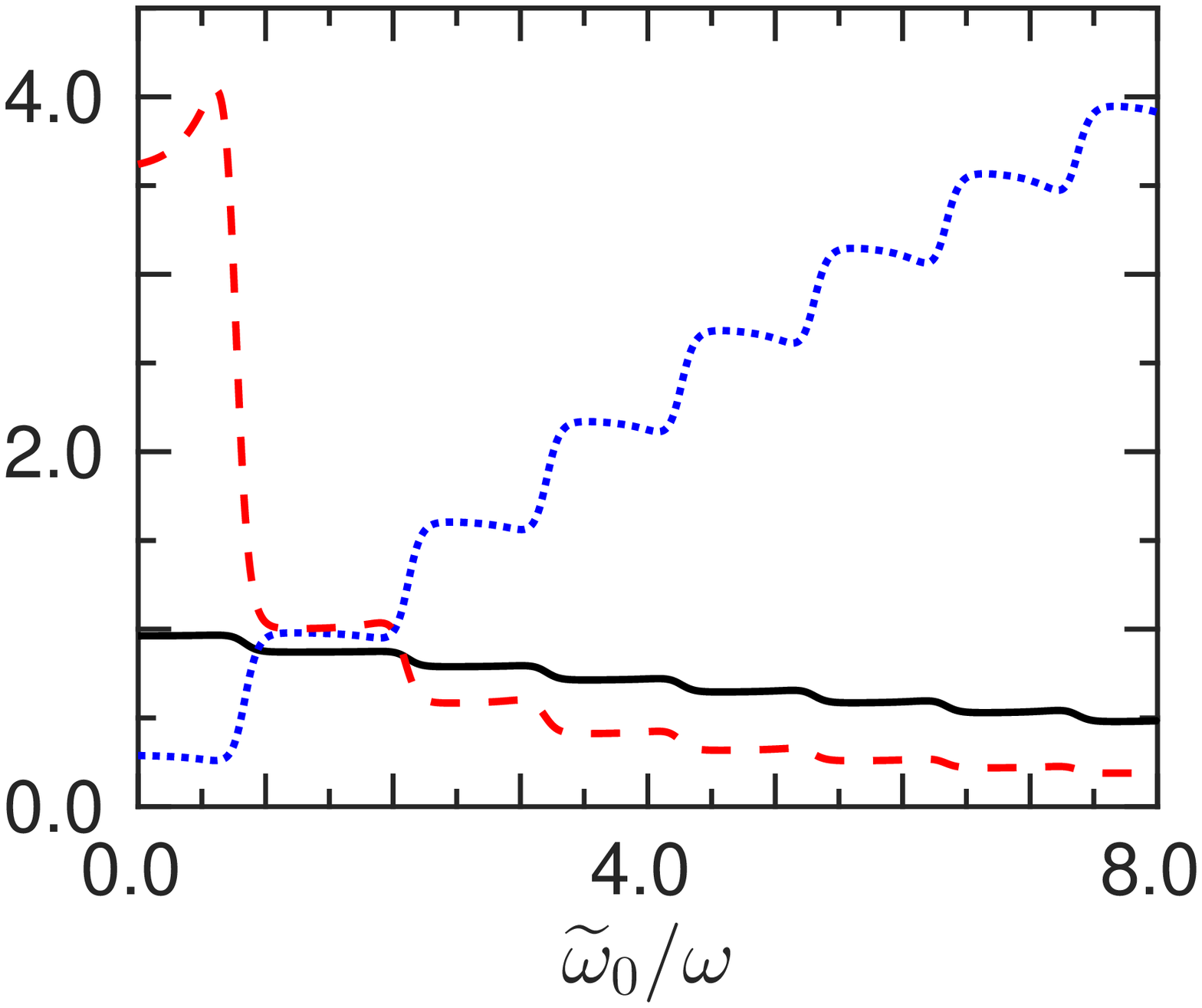}
\caption{As Fig.~\ref{F_1}, but for reduced width $\Delta\wo/\omega = 0.316$. 
	The sequence of plateaus is well captured by Eq.~(\ref{eq:APR}) with 
	$\ell_1 = -1,0,1,2,\ldots\;$.}  	   
\label{F_2}
\end{figure}

This changes when the density width is reduced to $\Delta\wo/\omega = 0.316$, 
while all other parameters are left unchanged, as shown in Fig.~\ref{F_2}: 
Here the scaled $n\!=\!0$-population features a series of well-developed 
plateaus with increasing center frequency which are explained with remarkable 
accuracy by Eq.~(\ref{eq:APR}), recalling $\nu/\omega \approx 1.387$ and 
$\beta\hbar\omega = 0.1$.

\begin{figure}[t]
\centering
\includegraphics[width=0.9\linewidth]{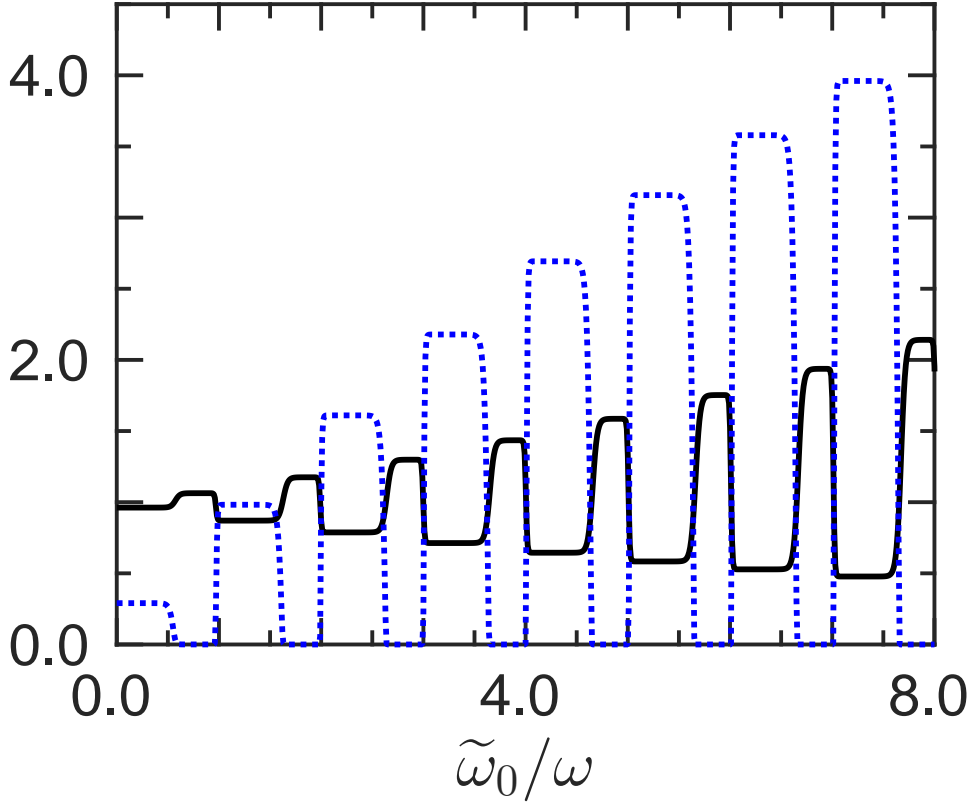}
\caption{Ratio~$r$ (full line; black) and scaled occupation probability 
	$p_0/P_0$ (dotted line; blue) of the Floquet state $n = 0$. All 
	parameters are identical to those employed in Figs.~\ref{F_1} 
	and~\ref{F_2}, except for the density width which is reduced further 
	to $\Delta\wo/\omega = 0.1$ here. Plateau values of $p_0/P_0$ with 
	$r < 1$ are governed by Eq.~(\ref{eq:APR}) with 
	$\ell_1 = -1,0,\ldots, 6$; intervals of quasithermal instability with 
	$r > 1$ by Eq.~(\ref{eq:BAD}) with $\ell_2 = -2,-3,\ldots,-9$. Observe 
	that $p_0/P_0 \approx 4$ for $7 \lesssim \wo_0/\omega \lesssim 7.5$, 
	corresponding to $\tau/T_{\rm bath} \approx 0.19$.}    
\label{F_3}
\end{figure}

A noteworthy phenomenon occurs when the density peak is made still sharper: 
Figure~\ref{F_3} depicts both $r$ and $p_0/P_0$ for $\Delta\wo/\omega = 0.1$. 
Now one observes {\em two\/} plateau sequences, intervals of successively 
lower~$r$ and, hence, lower quasitemperatures allowing for higher 
$n\!=\!0$-population alternating with intervals of successively higher~$r$, 
indicating successively stronger quasithermal instability. For explaining this 
numerical finding one has to go back to the  exact Eq.~(\ref{eq:RAT}), and to 
appreciate the fact that only the absolute value of the Floquet transition 
frequencies enters into the density~$J(|\wo|)$. Thus, when a peaked density 
such as described by Eq.~(\ref{eq:GSD}) is appreciably large at a positive 
frequency $\nu + \ell_1\omega$, it may also be large at a {\em negative\/} 
frequency $\nu + \ell_2\omega$, then leading to      
\begin{equation}
	r \approx \frac{N(-\nu - \ell_2\omega) + 1}{N(-\nu - \ell_2\omega)} 
	= \exp\!\big[ - \beta\hbar(\nu + \ell_2\omega)\big] > 1 \; . 
\label{eq:BAD}	 
\end{equation} 
This is the key to understanding Fig.~\ref{F_3}: The sequence of stable 
plateaus with $r < 1$ again is explained by Eq.~(\ref{eq:APR}) with 
$\ell_1 = -1,0,\ldots 6$ to an accuracy on the sub-percent level, while the
zones of quasithermal instability are governed by this Eq.~(\ref{eq:BAD}) with
$\ell_2 = -2,-3,\ldots,-9$. In principle, both kinds of processes have been 
competing already in the situation considered in Fig.~\ref{F_2}, but there 
the ``bad'' processes have been overshadowed because of their smaller Fourier 
weights. It is only the narrow peak width employed in Fig.~\ref{F_3} which 
allows one to disentangle the ``good'' transitions from the ``bad'' ones. 
It deserves to be pointed out that one reaches a scaled $n\!=\!0$-population 
$p_0/P_0 \approx 4$ for $7 \lesssim \wo_0/\omega \lesssim 7.5$, as 
corresponding to the scaled quasi\-temperature $\tau/T_{\rm bath} \approx 
0.19$. Thus, our present proof-of-principle study vindicates that Floquet-state
cooling can be fairly effective. Among others, our results imply that an ideal
Bose gas, stored in a parametrically driven oscillator trap, can condense
into a macroscopically occupied single-particle Floquet state even if the
ambient temperature is higher than the usual critical temperature. We also
point out that that this novel type of ``cooling by driving'' is 
thermodynamically consistent: The non-equilibrium steady state is characterized
by an energy flow which is always directed from the driven system into the 
bath, even if the quasitemperature of the former is lower than the actual 
temperature of the latter~\cite{DiermannEtAl19}.

\section{Discussion}
%%%%%%%%%%%%%%%%%%%%

The integrable model of the parametrically driven oscillator~(\ref{eq:HAM}) 
features an unusually simple quasienergy spectrum~(\ref{eq:QES}) in its 
stability regime. Combined with a system-bath coupling of the natural 
form~(\ref{eq:SBC}) it gives rise to a merely tridiagonal Floquet transition 
matrix~(\ref{eq:TOR}) and therefore leads to the expression~(\ref{eq:RAT}) 
which allows one to discuss the effect of the environment on the 
quasistationary state in an exceptionally transparent manner. Yet, the very 
essentials of the Floquet-state cooling mechanism --- a rich Fourier spectrum 
of the Floquet transition matrix elements~(\ref{eq:FTM}), the components of 
which may be addressed individually with suitable densities of states --- 
will be present also in realistic, non-integrable periodically driven systems 
with dense pure point quasienergy spectra, even if their quasi\-stationary
states can no longer be characterized in terms of a quasitemperature. Moreover,
for such chaotic systems it may no longer be feasible to assign meaningful 
quantum numbers to the Floquet states which depend on the parameters in a 
continuous manner, because their quasienergies exhibit a dense net of avoided 
crossings~\cite{HoneEtAl09}; in particular, it may no longer be feasible to 
identify a ``Floquet ground state'' by continuity. Nonetheless, it will still 
be possible to select some Floquet state of interest, and to guide population 
into that state by means of suitable bath densities of states. Therefore, the 
mechanism that has been exemplified with the help of the model~(\ref{eq:HAM}) 
is not restricted to that model, but fairly general. For these reasons we 
expect that Floquet-state cooling may find practical applications, {\em e.g.\/},
with periodically driven solid-state systems interacting with a phonon bath 
predominantly at certain well-defined frequencies. One may also envision 
deliberate quasithermal engineering, amounting to the design of either 
favorable system environments, or of particular driving forms in order to 
enrich the Floquet system's Fourier content.

The theoretical framework employed here, based on the golden rule-type 
rates~(\ref{eq:PAR}), is equivalent to the standard Born-Markov 
approximation~\cite{BreuerPetruccione02}, but a parametrically driven harmonic 
oscillator coupled to $N$ bath oscillators constitutes an integrable system 
for any~$N$~\cite{HagedornEtAl86}. Thus, it will be worthwhile to explore 
whether and how the findings reported in the present matter-of-principle 
study are recovered in the proper limit $N \to \infty$ without invoking any 
approximation at all.

It is known that in many situations a periodic drive can be switched off 
adiabatically, such that the occupation probabilities of the Floquet states
remain almost constant, even if the switch-off takes place within only a few
driving cycles~\cite{BreuerHolthaus89,DreseHolthaus99}. On the other hand,
the time scales required for relaxing to the quasistationary distribution
of Floquet-state occupation probabilities depend on the coupling to the 
bath which is weak by assumption. Therefore, under appropriate conditions it 
should be feasible to switch off the drive in an effectively adiabatic manner 
on times scales significantly shorter than the relaxation times after the
quasistationary state has been reached, so that its Floquet-state occupation 
probabilities determine the occupation probabilities of the system's proper 
energy eigenstates. This sketch might yield a blueprint for an actual cooling 
mechanism, providing higher-than-thermal ground-state populations.

\vspace{3ex}

{\bf DATA AVAILABILITY}

\vspace{1ex}
 
No datasets other than those plotted in Figs.~\ref{F_1} -- \ref{F_3}
were generated or analysed during the current study.

\vspace{5ex}

\begin{acknowledgments}
This work has been supported by the Deutsche For\-schungsgemeinschaft (DFG,
German Research Foundation) through Project No.~397122187. We wish to thank 
the members of the Research Unit FOR~2692 for many stimulating discussions. 
We are particularly grateful to Heinz-J\"urgen Schmidt and J\"urgen Schnack 
for insightful comments.
\end{acknowledgments}

\vspace{5ex}

{\bf AUTHOR CONTRIBUTIONS}

\vspace{1ex}

O.R.D.\ performed the numerical calculations, and contributed to the 
interpretation of the results. M.H.\ devised the concept, 
supervised the project, and wrote the initial version of the manuscript.
Both authors revised the manuscript.

\vspace{5ex}

{\bf ADDITIONAL INFORMATION}

\vspace{1ex}

The authors declare no competing interests.

\end{document}